\documentclass[conference]{IEEEtran}
\IEEEoverridecommandlockouts
\usepackage{cite}
\usepackage{amsmath,amssymb,amsfonts}
\usepackage{algorithmic}
\usepackage{graphicx}
\usepackage{textcomp}
\usepackage{xcolor}
\def\BibTeX{{\rm B\kern-.05em{\sc i\kern-.025em b}\kern-.08em
    T\kern-.1667em\lower.7ex\hbox{E}\kern-.125emX}}
 
\usepackage[numbers]{natbib}
\usepackage{xurl}
\usepackage{booktabs} 
\usepackage[normalem]{ulem}
\usepackage{xcolor}
\usepackage{color,soul}
\usepackage{xspace}
\usepackage[utf8x]{inputenc} 
\usepackage{textalpha}
\usepackage{url}
\usepackage{tikz}
\usetikzlibrary{arrows,shapes}
\usepackage[gen]{eurosym}
\usepackage{subcaption}

\usepackage{etoolbox}
\usepackage{xstring}
\usepackage{enumitem}
\usepackage{tikz}

\usepackage{etoolbox}
\usepackage{xstring}
\usepackage{enumitem}
\newcommand{\eg}{{\it e.g., }}
\newcommand{\ie}{{\it i.e., }}

\graphicspath{{figure/}{figures/}} 
\begin{document}





\title{Analysis of Independent Learning in Network Agents: A Packet Forwarding Use Case




}

\author{\IEEEauthorblockN{Abu Saleh Md Tayeen}
\IEEEauthorblockA{\textit{Dept. of Computer Science} \\
\textit{New Mexico State University}\\
Las Cruces, NM, U.S.A. \\
tayeen@nmsu.edu}
\and
\IEEEauthorblockN{Milan Biswal}
\IEEEauthorblockA{\textit{Dept. of Computer Science} \\
\textit{New Mexico State University}\\
Las Cruces, NM, U.S.A. \\
milanb@nmsu.edu}
\and
\IEEEauthorblockN{Abderrahmen Mtibaa}
\IEEEauthorblockA{\textit{Dept. of Computer Science} \\
\textit{University of Missouri--St Louis}\\
St. Louis, Missouri, U.S.A. \\
amtibaa@umsl.edu}
\and
\IEEEauthorblockN{Satyajayant Misra}
\IEEEauthorblockA{\textit{Dept. of Computer Science} \\
\textit{New Mexico State University}\\
Las Cruces, NM, U.S.A. \\
misra@cs.nmsu.edu}
\and

}

\maketitle

\begin{abstract}
Multi-Agent Reinforcement Learning (MARL) is nowadays widely used to solve real-world and complex decisions in various domains. While MARL can be categorized into independent and cooperative approaches, we consider the independent approach as a simple, more scalable, and less costly method for large-scale distributed systems, such as network packet forwarding. 
In this paper, we quantitatively and qualitatively assess the benefits of leveraging such independent agents learning approach, in particular IQL-based algorithm, for packet forwarding in computer networking, using the Named Data Networking (NDN) architecture as a driving example. We put multiple IQL-based forwarding strategies (IDQF) to the test and compare their performances against very basic forwarding schemes and simple topologies/traffic models to highlight major challenges and issues. 
We discuss the main issues related to the poor performance of IDQF, and quantify the impact of these issues on isolation when training and testing the IDQF models under different model tuning parameters and network topologies/characteristics.   
\end{abstract}

\begin{IEEEkeywords}
NDN Forwarding, Reinforcement Learning, Multi-agents, Independent Learning, Forwarding Strategy
\end{IEEEkeywords}

\section{Introduction}
\label{sec:intro}
Reinforcement learning (RL)~\cite{sutton1998introduction} helps explore and interact with local environment to acquire knowledge to solve real-world problems, especially complex multi-agent systems such as traffic signal control~\cite{wei2019presslight}, controlling drones~\cite{hung2016q}, robot soccer~\cite{schwab2018zero}, machines in factories~\cite{kazmi2019multi}, and automated trading~\cite{jin2018real}. RL-based solutions often deploy (i) a single centralized agent~\cite{stampa2017deep}, or (ii) multiple cooperative agents~\cite{gupta2017cooperative} to acquire the knowledge of such distributed systems. While these solutions may not scale for large distributed systems due to the amount of data exchange (\ie overhead), and additional command and control delays, we investigate in this paper the efficacy of non-cooperative/autonomous agents approach, Independent Q-Learning (IQL)~\cite{tan1993multi, wei2019presslight} and quantitatively and qualitatively assess its performance 
using packet forwarding in an ISP-scale network as a use-case.

ISP networks are often large-scale and complex, which renders using (i) a single centralized agent (\eg SDN), or (ii) multiple cooperative agents (\eg message exchange among routers) very costly--large overhead, single/multiple point of failure, and hard to maintain/upgrade. 
We assess the potential benefits and efficacy of using independent agents that learn to make forwarding decisions using only localized information (\eg incoming, outgoing traffic state). We choose to implement all our IQL-based models using Named Data Networking (NDN), which provides many ``useful'' features such as an easy to use forwarding plane, and context awareness decisions at the routers, to test our models, configurations, and tuning parameters. We provide a  short NDN primer and a motivation behind our NDN choice in Section~\ref{sec:motiv}. However, we believe that the paper's discussions, outcomes, and conclusions remain valid for IP architecture networks.

We compare the performance of a neural network equipped IQL-based forwarding strategy, IDQF (Independent  Deep  Q-Network Forwarding) against the most basic (\ie the simplest) forwarding scheme in NDN, namely {\em best\_route} (BR)~\cite{yi2013case}. We show that BR outperforms IDQF in most of the scenarios by simply choosing one path, while IDQF often fails to select the most appropriate path towards the producer.  

This comparison mainly aims at highlighting the challenges of IQL-based approach in such environment. We ran multiple experiments with various parameters, tuning, and setups to outline the challenges that led to unsuccessful decisions made by IDQF. The identified challenges (not a complete list) include: (i) the non-stationarity of the system due to the lack of global synchronization leading to triggered changes as soon as one agent changes its policy; (ii) the generalization of the training setup that is difficult due to network delays and frequent changes in the environment; (iii) the inability to assess the reward of an action immediately due to the variations of network delays; and 
(iv) the partial observability that prevents the routers to see overlapping paths, conflicting flows, etc. and make optimal decisions.
\textcolor{black}{We discuss these challenges in detail and present experiments and/or examples to support the impact of these challenges on the performance of  IDQF in particular, and distributed autonomous multi-agent RL schemes in general.}

The rest of this paper is organized as follows. Section~\ref{sec:motiv}
includes background and related works for this work. Section~\ref{sec:RL} summarizes the design of the IDQF forwarding mechanism and describe the simulation results. In Section~\ref{sec:challenge}, \textcolor{black}{we discuss the challenges regarding the IDQF forwarding scheme and autonomous multi-agent RL schemes.} Finally, Section~\ref{sec:conclusion} concludes the paper and suggest future work.

\section{Background And Related Work}
\label{sec:motiv}

\subsection{ICN/NDN Background}
\label{subsec:ndn}
Different from IP networks that use IP addresses (hosts identities) to inform intermediate routers' forwarding decisions, the fundamental idea of Named Data Networking (NDN) architecture is to retrieve the named-pieces of information (named network-layer packets), from any node that can provide it. 
In NDN, routers are equipped with a content store (CS), a pending interest table (PIT), and a forwarding information base (FIB). 
The FIB (similar to the forwarding table in IP routers) 
gets populated using a routing algorithm. 
Any node/router that receives a special packet (known as Interest packet) requesting for a content (Data packet) 
performs a CS-lookup on the content name. If the content is not available in the CS, the router performs a lookup in its PIT to check whether there is an existing request for the content that the router has forwarded into the network. 
If the PIT lookup is successful, the router adds the incoming Interest packet's interface to the PIT entry (interest aggregation) and drops the Interest. 
If no PIT match is found, the router creates a new PIT entry for the Interest. It then utilizes a forwarding strategy to select the best interface to upstream router(s) from multiple interfaces obtained from the FIB and forwards the Interest packet in the direction of the data source(s). 

\subsection{Related Work}
There exists a good number of literature which applied RL to routing in networks~\cite{mammeri2019reinforcement}. We mainly outline a few research works that applied RL for designing forwarding strategies in NDN. 
In~\cite{chiocchetti2013inform,zhang2020afsndn}, authors proposed Q-routing based strategies that alternates between exploration and exploitation phases to learn the best interface in terms of a cost (\textit{e.g.} delay) during dynamic network conditions. However, the switching of the phases are dependent on some parameters and hampers the ability of the strategies to react quickly to network changes such as congestion. Moreover, it requires altering the default NDN packet formats and data structures to calculate experience information for Q-routing, which violates the principle of
data immutability in NDN. 
The works~\cite{zhang2018ifs} and~\cite{de2020dqn} exploited neural networks to learn the mapping from the state of the network condition to the action of choosing the optimal interface for forwarding interests. However, there is no balancing of load across interfaces. Moreover, the strategies were verified using a single agent topology, which makes it difficult to assess the efficacy in larger/real networks.  

\section{IQL-based Stateful Forwarding}

\label{sec:RL}
In this section, we illustrate our IQL-based NDN forwarding scheme, IDQF where each node makes autonomous forwarding decisions based on their own observations and rewards. We also assess the performance of this scheme by comparing it to 
the {\em best\_route} (BR)~\cite{yi2013case}.





\subsection{Design of IDQF Forwarding Scheme}
\begin{figure}[htbp]
  \centering
  \includegraphics[width=0.99\linewidth]{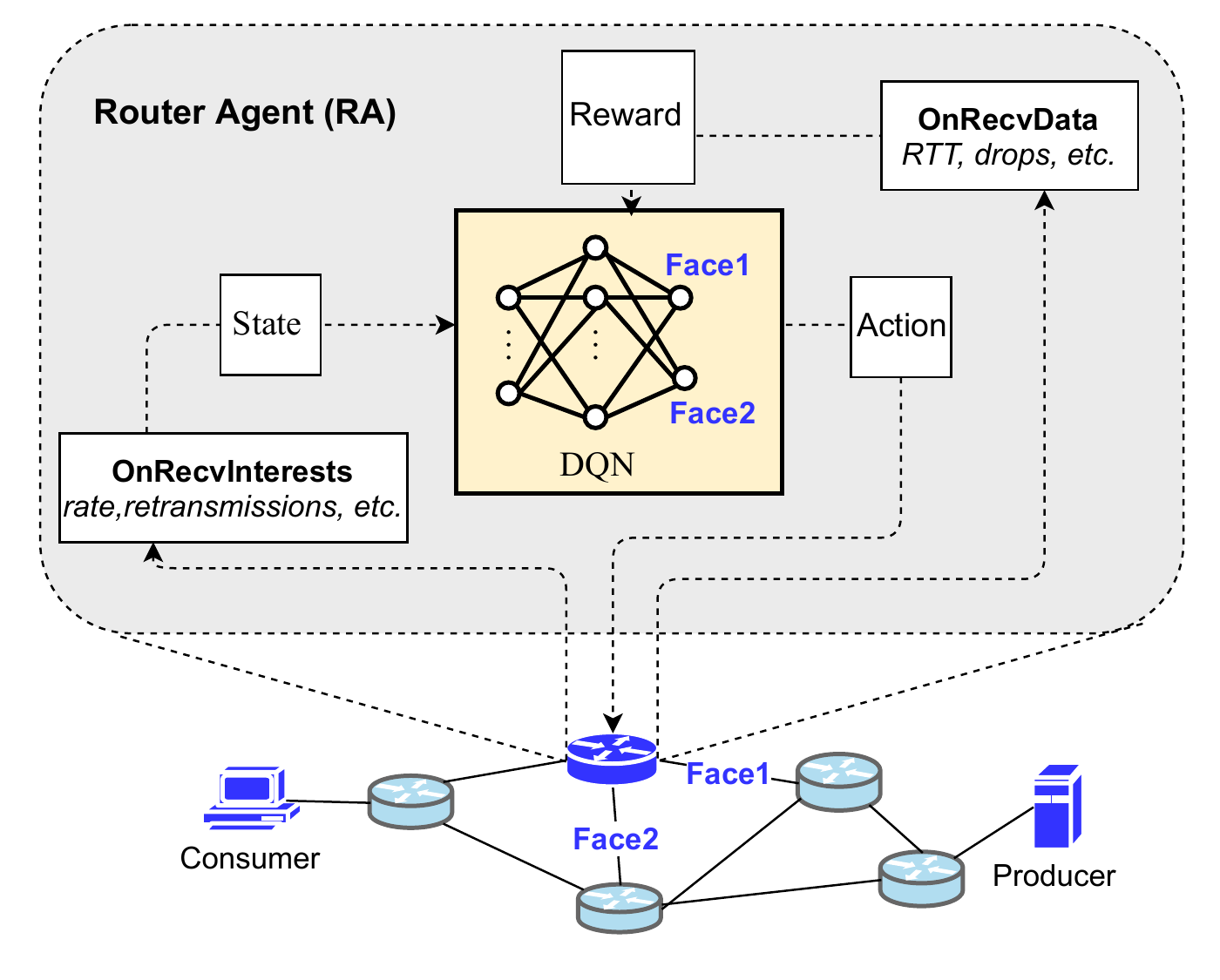}
  \caption{High level design of the IDQF forwarding strategy implemented at any router in the network. Periodically, routers update their states when receiving incoming Interests, and measure rewards when Data packets are received from the producer.  
  }
  \label{fig:rl-model-2}
\end{figure}

The goal of the agent in each router is to autonomously learn a decision policy to forward Interest packets through the optimal next-hop interface. The schematics of the IDQF forwarding scheme
is shown in Figure~\ref{fig:rl-model-2}. The agent learns/discovers the optimal forwarding policy indirectly using the action-value function. 
In the vanilla Q-learning approach, the state-action pairs are mapped to corresponding Q-values (maximum expected rewards) by maintaining a Q-table. For a communication network with numerous routers and communication flows, the size and complexity of state-action pairs or the corresponding Q-table will be very large, which poses a significant challenge for IQL to learn the optimal decision policy. To resolve this, we  approximate the action-value function using neural networks and use the Deep-Q-Network (DQN) training algorithm~\cite{mnih2015human}.

As shown in Figure~\ref{fig:rl-model-2}, the router $RA$ is equipped with neural network DQN.
At decision time $t$, $RA$, calculates the state of its available next hop interfaces, \textit{Face1} and \textit{Face2} by observing some local parameters such as round trip times (RTTs), interest satisfaction ratios, number of re-transmitted packets collected between time $t$ and $t-\delta t$. Here, $\delta t$ is the time period for sensing and updating the weights of the DQN. The state is fed into the DQN as input. DQN then predicts either \textit{Face1} or \textit{Face2} as an action. The $RA$ agent then forwards the current Interest packet and all subsequent Interest packets through the predicted interface until next decision time occurs at $t+\delta t$. The chosen action results in a change of state for the corresponding interface. $RA$ also gets a feedback for its action, which is the reward. We define the reward, $RW(t)$ at time $t$ by the following equation: 
\begin{equation}
\label{eq:rw}
    RW(t) = -\Big[ \Big( \frac{1}{M}\sum_{k=(t-\delta t)}^t{RTT(k)} \Big) + C \times R(t) \Big]
\end{equation}
Here, $M$ is the number of Data packets received between $t$ and $t-\delta t$, 
$RTT(k)$ is the Round-Trip Time (RTT) of the received Data packet at time $k$, $R(t)$ is the number of re-transmitted packets which were actually transmitted through the interface chosen for the $\delta t$ period, and $C$ is a penalty constant multiplier that aims at discouraging choosing interface with high drop ratios (we choose $C=4$ seconds in our case). Note that in case of drops of Interests/Data packets, there will be no RTT measurements, thus the penalty constant helps in assigning a maximum delay for every packet dropped. 
The $RA$ agent uses the reward to improve its decision policy.

\subsection{Evaluation of IDQF}
\begin{figure}[htbp]
  \centering
  \includegraphics[width=0.9\linewidth]{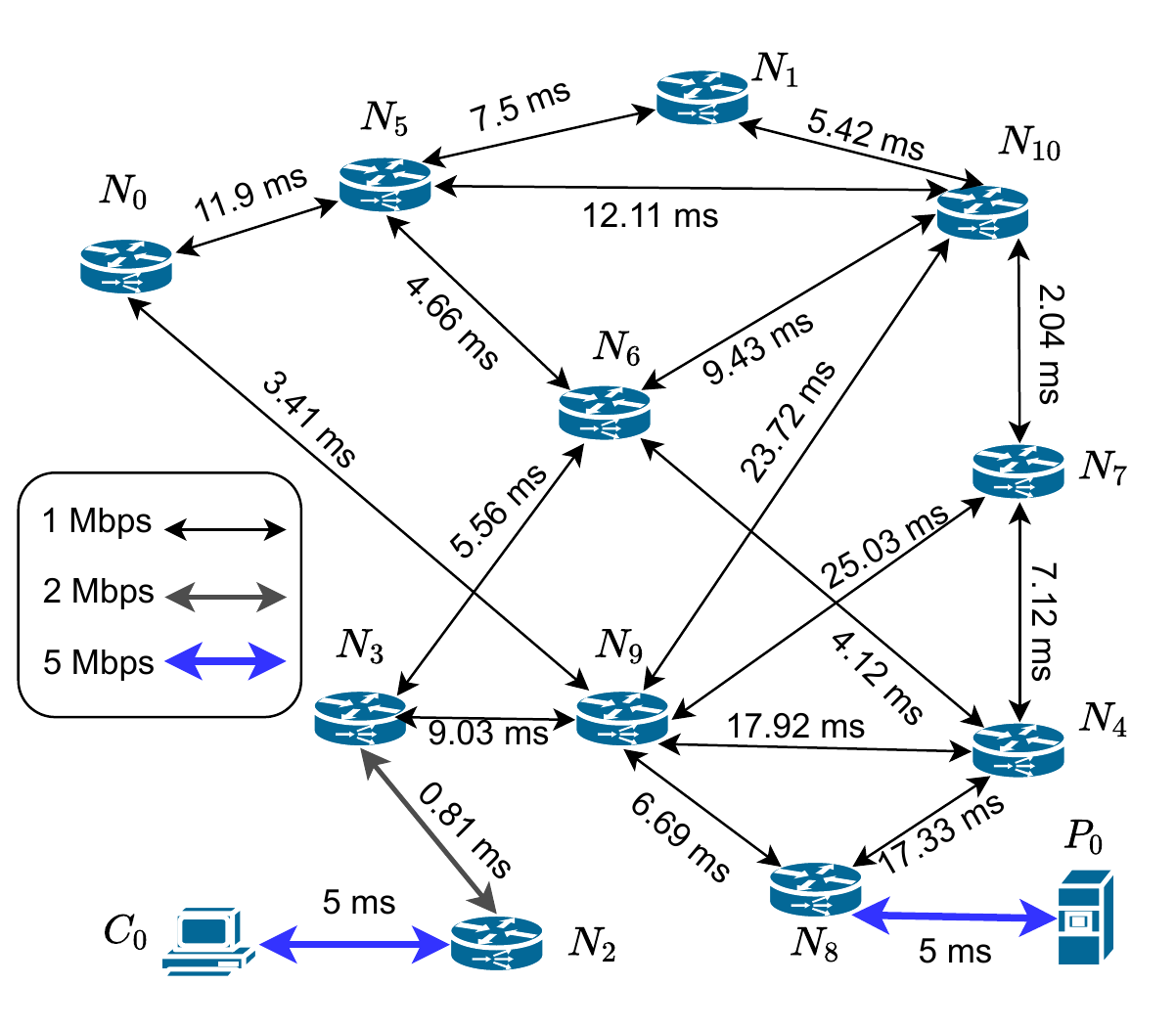}
  \caption{Sprint topology used in our simulation; propagation delays (in milliseconds) are shown as link weights; link capacities vary from 1Mbps at the core (black links), 2Mbps (grey) and 5Mbps at the edge (blue)} 
  \label{fig:topo-Sprint-2}
\end{figure}

We use \emph{ndnSIM}~\cite{mastorakis2017ndnsim}, the NDN network simulator, to evaluate the forwarding scheme performance. 
To implement the RL agents, 
we used a concurrently running daemon process that interfaces with the \emph{ndnSIM} and exchanges training data and prediction of the next hop interface for each node. In this paper, we will share the results using the Sprint topology shown in Figure~\ref{fig:topo-Sprint-2}. We have  tested the performance using multiple network topologies (grid, tree), and performed tuning of different parameters of DQN.


We consider a simple scenario, where only one consumer ($C_0$) requests data from only one producer ($P_0$) at a fixed rate $r$. Note that we used such a simple scenario to highlight challenges that may arise even in a very predictable network traffic with no loss, errors, contention, etc. 
The Sprint topology in Figure~\ref{fig:topo-Sprint-2} consists of $11$ router nodes and $18$ links among them. 
Each router node has a queue size of 100 packets. 
In our simple scenario, we deploy the agents in nodes $N_3$, $N_4$, $N_6$, and $N_9$ only. Each of these agents have their own IDQF-strategy.  
In our results, we have considered the size of the state vector to be two and chose the best two interfaces at each agent (those with the least delays to the producer). Note that we have used and tested different sizes which result in performance degradation when we scale the exploration beyond two interfaces. 

\begin{figure}[!th]
\begin{subfigure}{0.46\columnwidth}
\centering
\includegraphics[width=\linewidth]{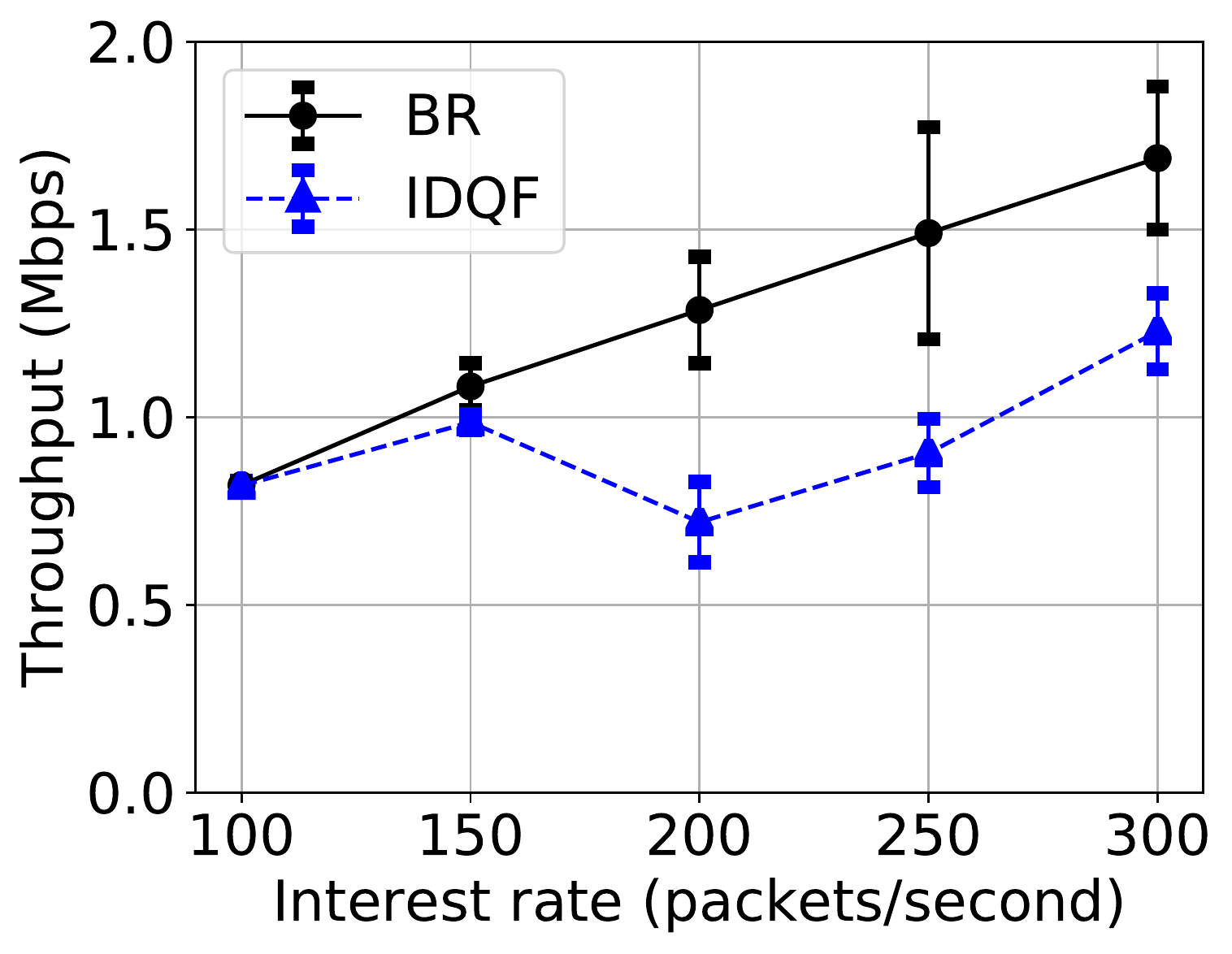}
\caption{Throughput measurements at the consumer node $C_0$}
\label{fig:thrpt-comp}
\end{subfigure}
\begin{subfigure}{0.46\columnwidth}
\centering
\includegraphics[width=\linewidth]{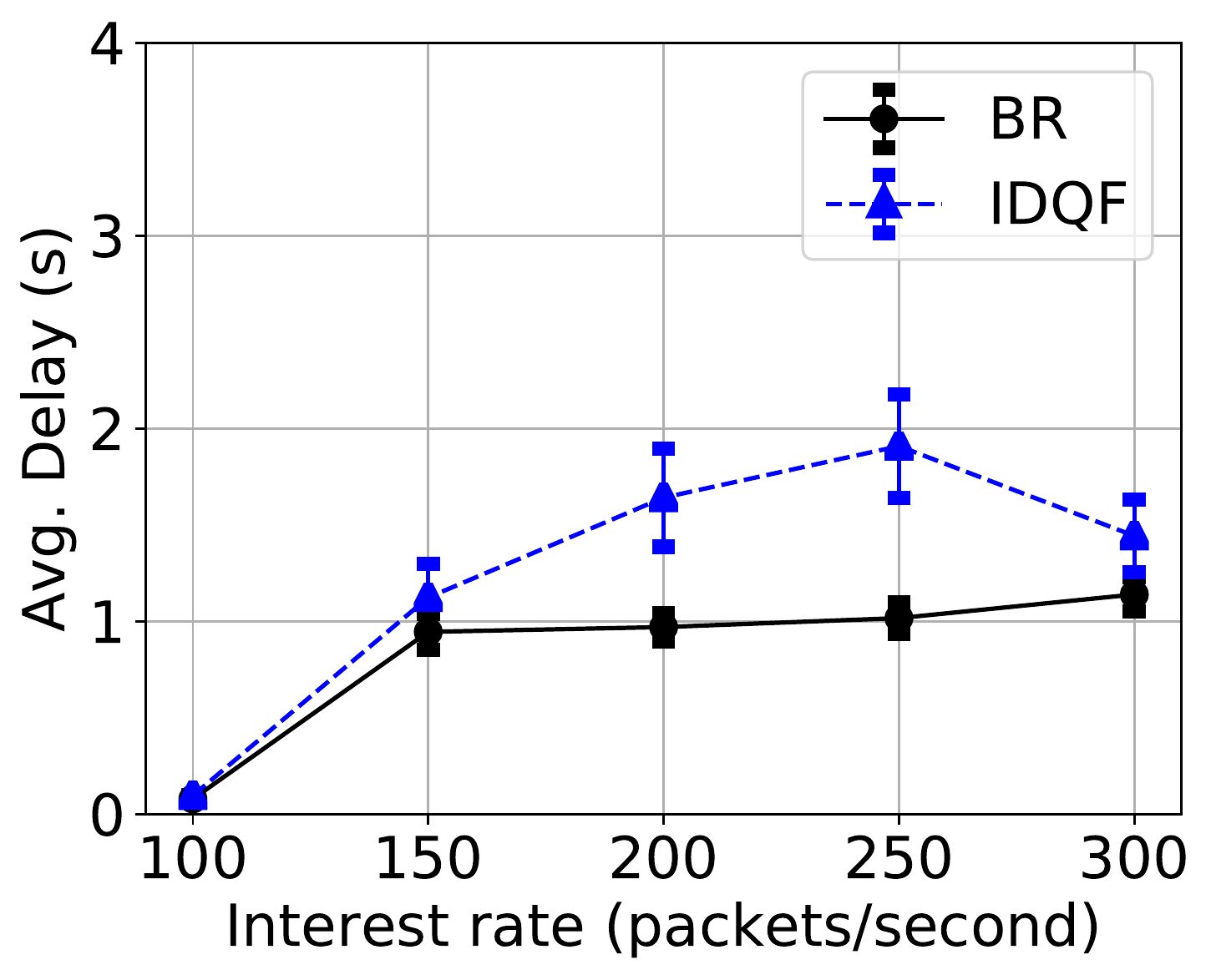}
\caption{End-to-end delay measurements at the consumer node $C_0$}
\label{fig:delay-comp}
\end{subfigure}
\caption{Throughput and delay comparison of \emph{best\_route} (BR) and IDQF performance; Error bars indicate the variation of the measurements during the simulation} 
\label{fig:thrpt-del-comp}
\end{figure}

In our simulations, the DQN of each agent has one hidden layer, one input layer and one output layer. Each hidden layer
consists of 32 nodes, and the layers are fully connected with
each other.
While training the node agents we tuned several parameters of DQN. The first parameter is the decay rate, which controls the number of times the agents will perform exploration by randomly choosing actions. Second, we adjusted the learning rate to control the convergence speed of DQN. Third, we varied the capacity of the replay buffer which governs the diversity of samples available to the agent at any moment. 
We trained the DQN of agents with decay rate=$0.001$ and learning rate=$1.0$ for 50 episodes where an episode corresponds to one instance of simulation of the network environment with duration 60 seconds. The agents were trained separately for five different request rate starting from low Interest rate (41\% of network capacity) to high Interest rate (123\% of network capacity) in packets/second (pkts/s). The interest rates are 100 pkts/s (0.82 Mbps), 150 pkts/s (1.23 Mbps), 200 pkts/s (1.64 Mbps), 250 pkts/s (2.05 Mbps), and 300 pkts/s (2.46 Mbps). We deployed the agents for each Interest rate and ran simulations five times to get average results.
We compare the throughput and average delay measured in $C_0$ for both BR and IDQF scheme and present the results in Figure~\ref{fig:thrpt-del-comp}. 
The height of the bar for each point in Figure~\ref{fig:thrpt-del-comp} represents the standard deviation of throughput and delay measurements at every second of simulation.  

Figure~\ref{fig:thrpt-del-comp} shows that the IDQF strategy does not outperform even the most basic 
strategy, {\em best\_route} (BR). BR always uses the interface with the shortest delay regardless of congestion. It will react and switch an interface only if the best route fails (after at least one timeout event). Note that we share the IDQF configuration that provides the best results for this scenario. 
As shown in Figure~\ref{fig:thrpt-comp}, BR outperforms IDQF by 10\%-36\% in terms of throughput for all Interest rates. We notice that IDQF often fails to choose the best interface, provided the cost (\eg latency from the router to the producer) of using each interface. 
Similarly, in Figure~\ref{fig:delay-comp}, we find that IDQF strategy fails to deliver data faster than BR in most cases, especially at lower Interest rates. Interests are often forwarded to interfaces that lead to congestion and drops, resulting in an average of 2 to 3 re-transmissions per Interest. 

To summarize, IDQF strategy, while tested even with a simple scenario with a single flow and limited network variability, fails to converge into a stable state that learns the non-changing network characteristics and makes decisions that can often be solved by simple utility functions. 
In this study, we have identified a list of challenges (discussed in the next section) which advocate that reinforcement learning based approaches driven by autonomous decision making paradigm will result in similar non-optimal or sub-optimal solutions. These challenges are generalized beyond NDN, and may apply to IP and other network architectures/systems.

\section{Challenges of using autonomous agents in distributed systems}
\label{sec:challenge}
\textcolor{black}{We will build on the results shown in the previous section that highlight the sub-optimal performance of the IDQF strategy, and autonomous multi-agent RL algorithms in general.} In this section, we will discuss and analyze qualitatively and quantitatively few challenges that contribute to this ``low'' performance.


\subsection{Training and Generalization of Agents}
The exhaustiveness of the sample training set and the method of training play a significant role in proper generalization of the neural network of an agent. In a multi-agent setting, the goal of every agent is to maximize its cumulative reward or the expected return, 
which is simply the sum of all rewards the agent received starting from the initial state to a terminal state of the environment. An agent can reach a terminal state if certain score is obtained or the task is complete. 
To learn an optimized policy the agents are iteratively trained through multiple episodes where each episode is a sequence of states, actions, and rewards starting from an initial state to a terminal state. Often, it is difficult to determine a terminal state or a target score for the network environment. 
Therefore, we argue that setting the episode as a period of time, which is long enough to incorporate actions, and their rewards, will depend on the network topology and its characteristics (traffic, congestion). Based on the choice of episode length (simulation duration), there is a possibility of over-fitting or under-fitting of the neural networks of the agents, that can often lead to bad generalization and sub-optimal policy learning for the agents. 


To better illustrate this challenge, let us consider the scenario depicted in Figure~\ref{fig:topo-Sprint-2}. Agents closer to the consumer node would receive much more Interest packets, resulting in more data points, when compared to others closer to the producer node. Indeed, agents explore their interfaces by spreading received Interests across their next hop nodes, resulting in fewer number of Interests received by core routers which render learning more challenging at these nodes. 
For instance, nodes such as $N_4$ will not get enough number of Interests to build its training data within the short simulation duration. In our controlled experiments with a fixed network topology, we tested and tuned different simulation periods, yet failed to find optimal forwarding strategies for the nodes.

The generalization of the agents also depends on the size of a memory buffer called \emph{experience replay}~\cite{mnih2015human} that is used to store experience data (\ie state transition record as tuple $<$state, action, reward, next state$>$). This replay buffer helps the neural networks to break correlation in observation sequences and learn from more independently and identically distributed past experiences to stabilise the training. Without replay buffer, agents will update the weights of their neural networks based on only current state, action, and reward. However, sometimes state transition tuples generated at earlier stage of exploration constitute a good policy for the agent. These records need to be stored and can be utilized for training the neural network to achieve better generalization. During training with replay buffer, the agents update the weights of their neural networks or DQN by randomly sampling mini-batches of experience data from it. 

\begin{figure*}[tbp]
\begin{subfigure}{0.32\textwidth}
\centering
\includegraphics[width=\textwidth]{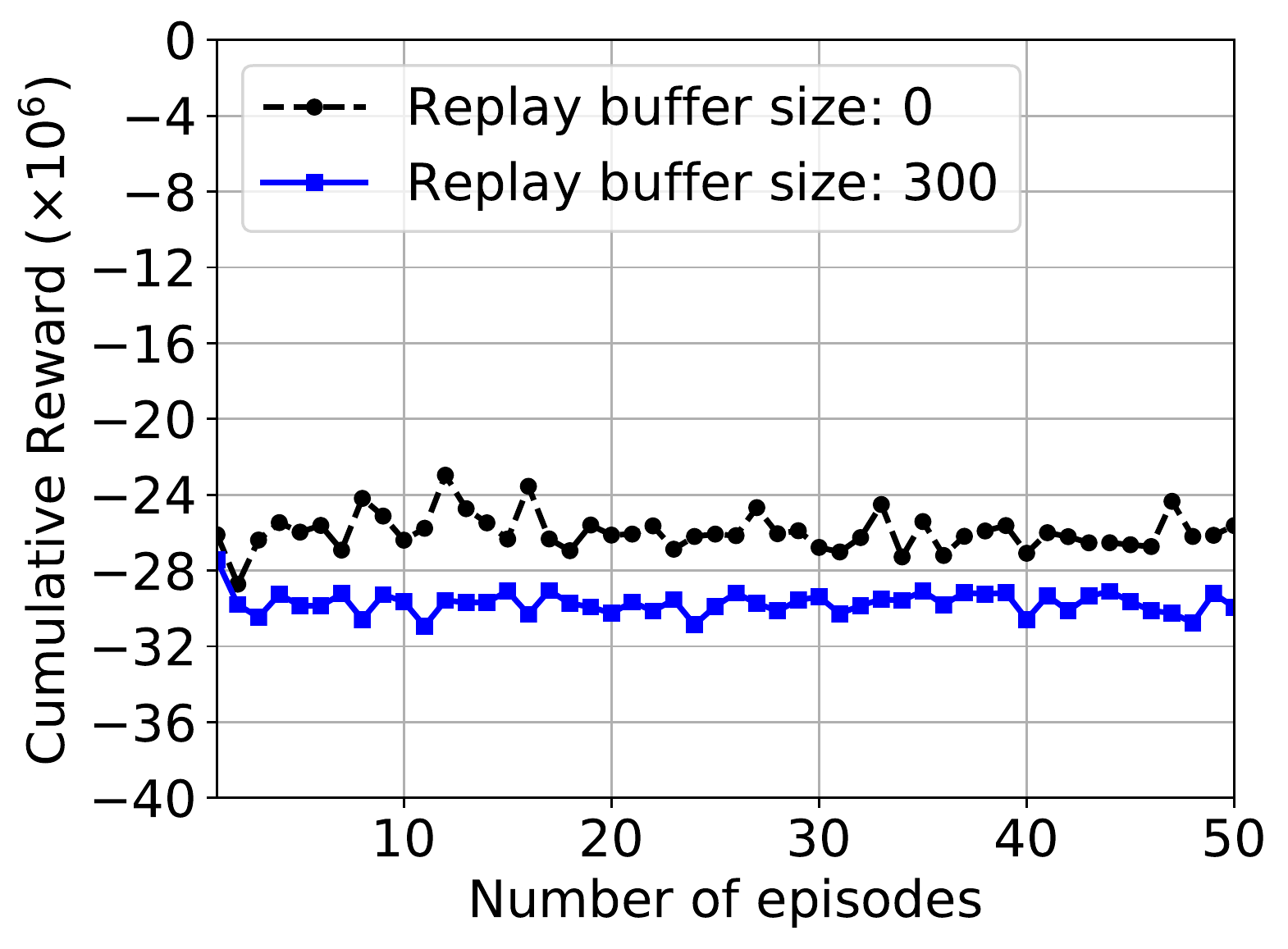}
\caption{Cumulative reward of node agent $N_3$ with different sizes of replay buffer}
\label{fig:reply-buff-N3}
\end{subfigure}
\hfil
\begin{subfigure}{0.32\textwidth}
\centering
\includegraphics[width=\textwidth]{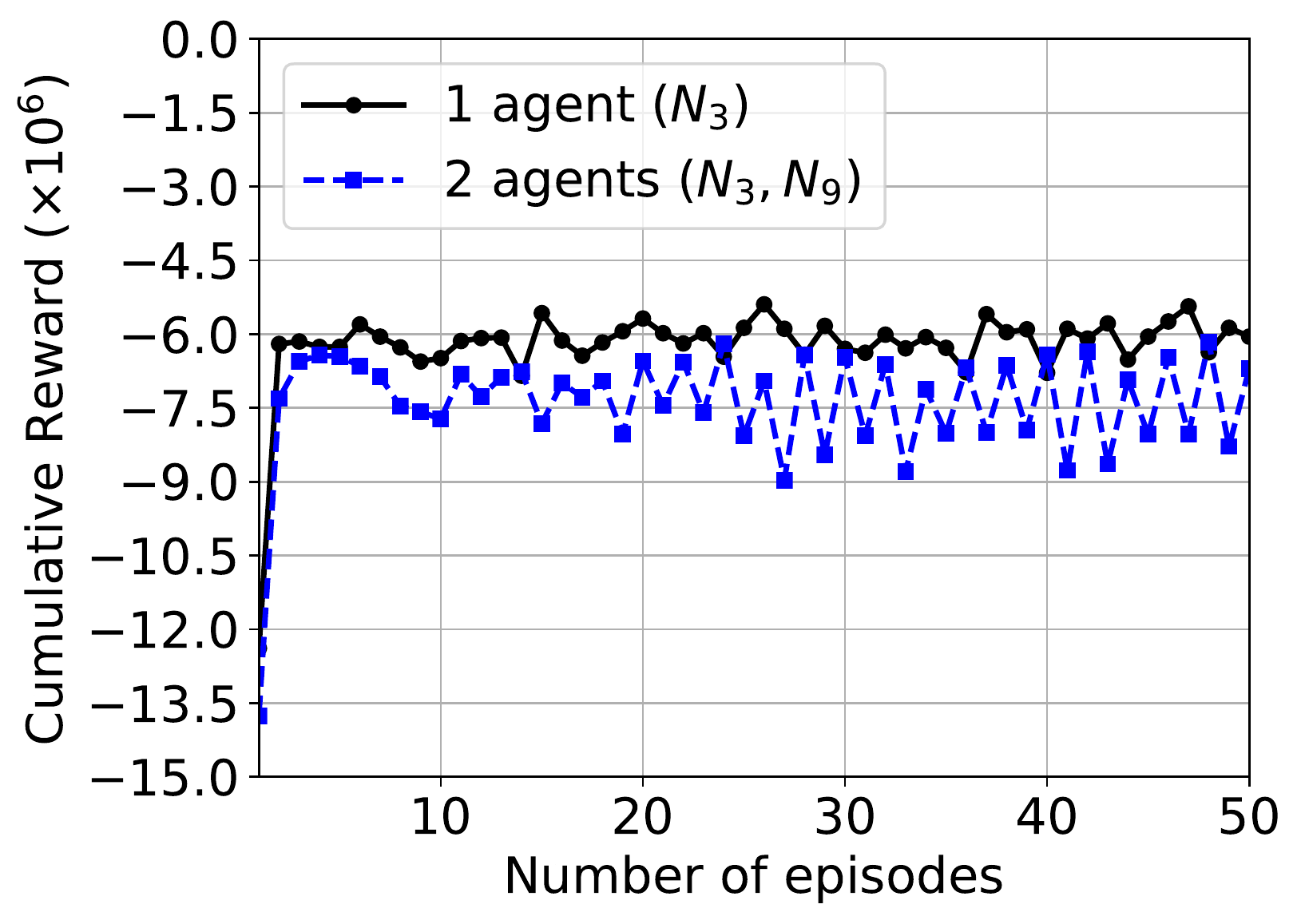}
\caption{Cumulative reward of node agent $N_3$ when trained with different number of agents}
\label{fig:nonstationary-N3}
\end{subfigure}
\hfil
\begin{subfigure}{0.32\textwidth}
\centering
\includegraphics[width=\textwidth]{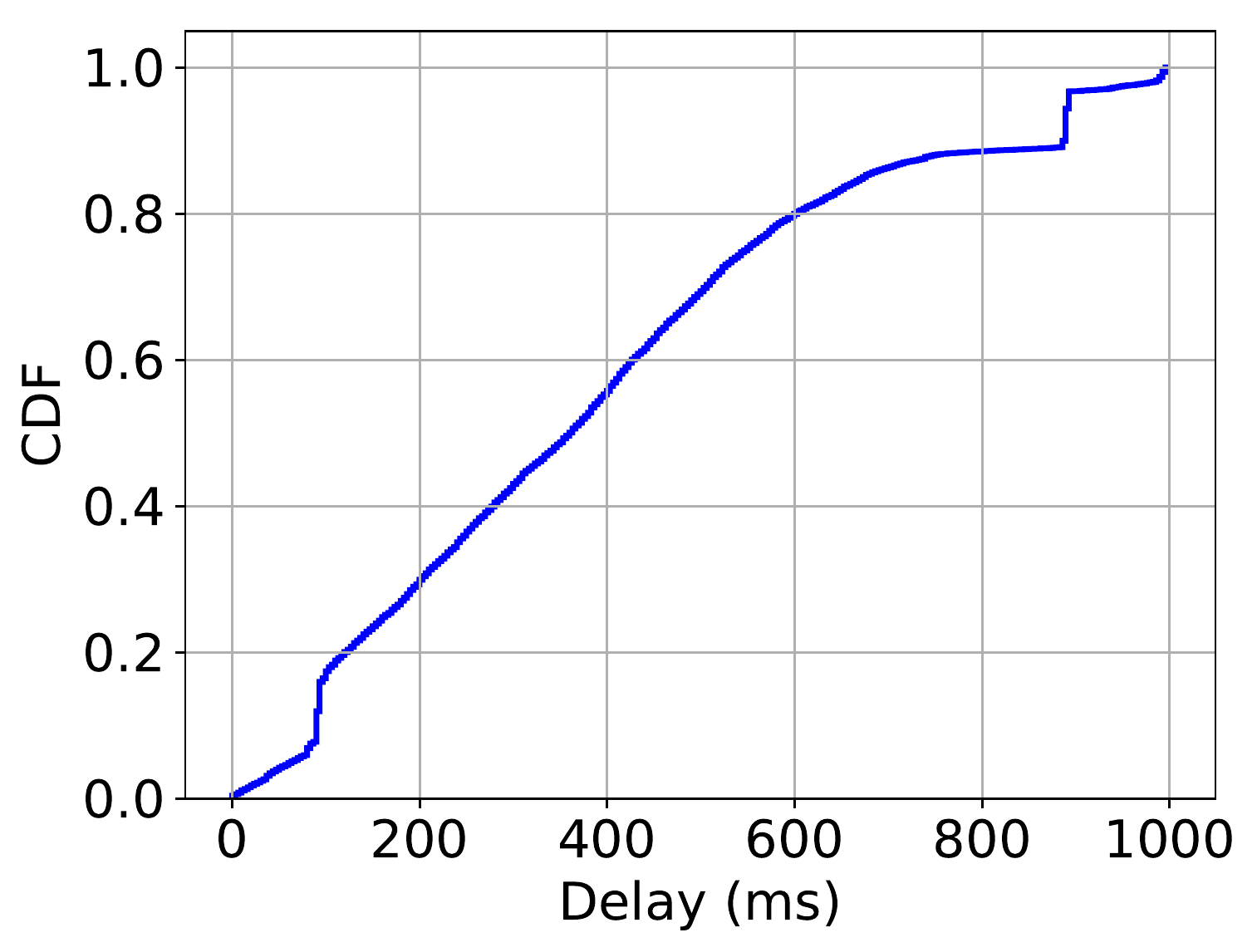}
\caption{CDF of delays of node agent $N_3$}
\label{fig:delay-cdf-N3}
\end{subfigure}

\caption{Demonstration of non-stationarity, effect of replay buffer size and delay measurements during training} 
\label{fig:rep-buff-non-stationarity}
\end{figure*}

To demonstrate the effect of replay buffer, we ran two experiments: one involving agents trained without any replay buffer and another by training agents with replay buffer. We show the cumulative rewards obtained by node agent $N_3$ per episode of training in Figure~\ref{fig:reply-buff-N3}.
We show that the absence of replay buffer leads to higher oscillation which makes the system more susceptible to variations in the input states resulting in issues in convergence across all agents in the network. Deploying a replay buffer (we have tested different buffer sizes) often exhibits less variations, however, at the same time the cumulative reward does not move toward higher values as training progresses through every episode. This results in additional delays in reacting to any sudden change in the network characteristic (\eg input state, reward). As a result, experience replay is difficult to tune in such autonomous multi-agent RL scenario.
\subsection{Non-stationarity of the Environment}

One of the major challenges for distributed autonomous multi-agent RL is the \emph{non-stationarity}. 
In a multi-agent setting of Markov Decision Process, the state and reward of each agent depend on the actions of all agents and every agent adapts its behaviors in the context of other co-adapting learning agents.
Since all agents are learning concurrently, the environment is not stationary and constantly changing from the perspective of any given agent. This phenomenon actually rules out any convergence guarantees.
To demonstrate the effect of non-stationarity we performed two experiments on topology in Figure~\ref{fig:topo-Sprint-2}. 
In one experiment, we put an agent only in node $N_3$ and train the single agent which will learn to choose between interfaces to node $N_6$ and $N_9$. In another experiment, we put agents in both $N_3$ and $N_9$ and train the two agents concurrently. Here, $N_9$ will learn to select between its next hops $N_8$ and $N_4$. We assume that in both cases all the other nodes without any agents know the forwarding next hop node to get optimal performance. 
In Figure~\ref{fig:nonstationary-N3}, we show the cumulative reward of node agent $N_3$ for the two cases. From Figure~\ref{fig:nonstationary-N3}, we observe that training with two agents cause the cumulative rewards of $N_3$ to be less and to fluctuate more compared to that of $N_3$ when training was done with one agent. 
This is because with two agents the input state and reward of agent $N_3$ depends on the actions (face selections) of agent $N_9$. Similarly, the input state of agent $N_9$ is conditioned upon $N_3$'s  distribution of interests to $N_9$. As a result, during learning both agents keep changing their policy of choosing an interface based on the actions taken by each other. These actions can be conflicting (as shown by the large oscillations in the two agent cumulative reward) and result in degradation of performance as shown in our previous results. On the other hand, with only one agent $N_3$, its state and rewards solely depend on its actions, resulting in faster convergence and better performance. This phenomenon is amplified when: (i)  all agents are running simultaneously, (ii) the number of agents in the network scales, (iii) the network experience multiple concurrent flows (\eg many consumers and producers). 
One of the ways to deal with non-stationarity is to adopt a centralized architecture where agents will know the actions of one another. However, it is infeasible for forwarding packets in large distributed computer networks.


    
    
\subsection{Partial Observability of the Environment}
Another key factor in autonomous multi-agent RL is the \emph{partial observability}. In many real-world environments, agents cannot perceive the complete state of the environment. For example, in our context of packet forwarding, the agents at each router observe the state per interface per producer in terms of some parameters such as RTT or drop ratio which indicate the congestion status of their respective interfaces. However, agents cannot distinguish certain events, which may occur beyond their local knowledge, including if two or more of their available interfaces lead to overlapping paths towards a given producer, or if a given interface will compete with other distant flows (\ie using portions of the path towards the producer). All these situations may result in similar measured rewards, but require different mitigation (\eg drop packets, choose different interface, etc.). Besides, the node agents neither know the actions (next hop choices) other agents have taken nor can compute the global reward (overall delay in consumer) by themselves. In order to find a globally optimal policy, the agents need to take into account the states and behavior of other agents and adapt to the joint behavior accordingly. Global observability, such as in SDN, or message-exchange-based coordination is needed to overcome this problem. We argue that even message-exchange-based coordination is challenging due to delays in message exchanged across agents unless a mechanism such as separate control channel is in place  to transmit the messages out of band. 
    
    
\subsection{Inability to Assess Actual Reward}
Most of the RL algorithms have been designed for environments where the feedback or reward is instantly available upon execution of an action. These algorithms cannot adequately solve the problems where actual rewards can not be assessed or are not  available immediately~\cite{shen2016reinforcement}. The delayed feedback phenomenon is also visible in our packet forwarding problem.
In any network, delays occur due to many factors including transmission, propagation, queuing, etc. Thus, measuring the state parameters and true reward value for the last action can not be collected before the next decision step occurs. Since RTTs could take from few milliseconds to seconds (especially due to drop and re-transmissions), estimating the current state and reward, \eg RTT or drops, for any forwarded Interest packet is pushed by additional and often variable delays. Throughout these waiting times, the agents will have to make forwarding decisions for incoming Interest packets and assign some approximated rewards against its state-action observations. These approximated rewards make it difficult to determine which of the state-action pairs contributed to the achievement of the optimization goal. This can lead to agents obtaining sub-optimal policies. 
For example, the topology in Figure~\ref{fig:topo-Sprint-2} shows that the theoretical RTT of the optimal path between router $N_3$ and the producer is 49.51ms.
However, as shown in Figure~\ref{fig:delay-cdf-N3}, based on our experiments we measured delays ranging from 61ms (\ie counting queuing and other processing delays) to 1s (maximum delay before dropping PIT entries), where more than 50\% of the delay measurements exceeded 300ms. 
These large variations in the delay measurements in a controlled experiment, where number of flows, traffic, and routes are static, pushes autonomous multi-agent RL approaches to make approximate forwarding decisions which are often far from the optimal policy.

\section{Conclusion}
\label{sec:conclusion}
In this paper, we applied IQL, the autonomous and non-cooperative multi-agent RL approach to perform packet forwarding in NDN. We approximated the Q-function of IQL using DQN. We performed many experiments in different topologies with tuning of several parameters to achieve a better generalization of the DQN. We show that our implemented forwarding scheme, IDQF does not significantly improve the network performance compared to the default forwarding strategy available in NDN. We have also presented several challenges (\eg non-stationarity, partial observability, etc.) of  this approach using examples and experiments to support its effect on the performance of IDQF. A possible direction for our future work is to try fully cooperative multi-agent reinforcement learning approaches to design forwarding strategies. 

\bibliographystyle{IEEEtran}
\bibliography{references}

\end{document}